\documentclass{article}
\usepackage{graphicx}
\usepackage{arxiv}
\usepackage{hyperref} 

\title{Data-driven prediction of structure of metal-organic frameworks}

\author{
    Elizaveta I. Yakovenko\\
    MSU Institute for Artificial Intelligence\\
    Lomonosov Moscow State University\\
    Moscow 119192, Russia\\
    \And
    \href{https://orcid.org/0000-0001-7008-4586}{\includegraphics[scale=0.07]{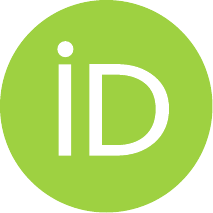}\hspace{1mm}Iurii M. Nevolin}\\
	Frumkin Institute of Physical Chemistry\\
	and Electrochemistry\\
    Russian Academy of Sciences\\
    Moscow 119071, Russia\\
    \And
    Anatoliy A. Chasovskikh\\
    Faculty of Mechanics and Mathematics\\
    Lomonosov Moscow State University\\
    Moscow 119991, Russia\\
    \And
    \href{http://orcid.org/0000-0001-8891-6862}{\includegraphics[scale=0.07]{orcid.pdf}\hspace{1mm}Artem A. Mitrofanov}\\
    Department of Chemistry\\
    Lomonosov Moscow State University\\
    Moscow 119991, Russia\\   
    \And
    \href{https://orcid.org/0000-0001-6117-5662}{\includegraphics[scale=0.07]{orcid.pdf}\hspace{1mm}Vadim V. Korolev}\thanks{\textit{Email address}: \texttt{V.Korolev@iai.msu.ru}}\\
    MSU Institute for Artificial Intelligence\\
    Lomonosov Moscow State University\\
    Moscow 119192, Russia\\
}

\begin{document}

\maketitle

\begin{abstract}
Crystal structure prediction (CSP) has proven to be a highly effective route for discovering new materials. Substantial advancements have been made in CSP of inorganic and molecular crystals, while hybrid materials, including metal–organic frameworks (MOFs), have been unfairly overlooked. The \textit{ab initio} techniques adopted for the CSP of MOFs cannot be scaled to a high-throughput regime, which is required for efficient exploration of the immense chemical space. Here, we propose a data-driven method to tackle current needs of computational MOF discovery. By examining CSP through the lens of reticular chemistry, coarse-grained neural networks were implemented to predict underlying net topology of crystal graphs. The models showed satisfactory performance, which was next enhanced by limiting the applicability domain. Flue gas separation was used as an illustrative example to validate the proposed CSP approach. Several hundred \textit{in silico}–generated systems revealed a notable discrepancy in adsorption capacity among competing polymorphs.
\end{abstract}

\section{Main}
Metal–organic frameworks (MOFs) are a class of potentially porous materials governed by reticular-chemistry principles\cite{yaghi2003reticular}. Assembly of molecular building blocks leads to extended crystalline structures, which are versatile platforms for adsorption-related applications\cite{moghadam2024progress}. Nonetheless, conventional methods for designing new functional MOFs, whether through modification of existing linkers or via substitution of secondary building units (SBUs), pose a notorious challenge: the lack of a straightforward relation between constituents and the resulting atomic configuration. The issue is resolved by reformulating it as a global optimization problem using computational approaches to crystal structure prediction (CSP)\cite{oganov2019structure}. Most studies related to CSP are focused on inorganic and molecular crystals, whereas framework materials, including MOFs, require special consideration. The number of atoms per unit cell amounts to hundreds, leading to a combinatorial explosion of configurational space\cite{oganov2006crystal}; the introduction of nonatomic building blocks can meaningfully reduce the set of trial structures\cite{mellot2004hybrid}. Furthermore, the rich polymorphism in pressure–temperature space\cite{widmer2019rich} necessitates advanced techniques to estimate the Helmholtz free energy of competing phases\cite{xu2023experimentally}; these techniques go beyond routine density functional theory (DFT).

The development of the latest \textit{ab initio} approaches to the CSP of MOFs involves incorporating domain-specific features into existing algorithms. The automated assembly of secondary building units method\cite{mellot2004hybrid}, initially designed for simulating inorganic frameworks\cite{mellot2000novo}, has been successfully adapted to the processing of both inorganic and organic counterparts as distinct building blocks. A more sophisticated technique\cite{darby2020ab} has been used to unify algorithms “\textit{ab initio} random structure searching”\cite{pickard2011ab} and “Wyckoff alignment of molecules”: symmetry constrains are imposed on random trial structures by placing molecules at special Wyckoff sites. The above methods are both intended to perform CSP of MOFs from first principles, i.e., without relying on \textit{prior} knowledge of experimental structures. Consequently, the universality of \textit{ab initio} algorithms comes at a cost of thorough exploration of configurational space, involving numerous DFT calculations of energy, forces, and stresses. In principle, machine learning interatomic potentials trained on diverse MOF data\cite{sriram2023open} can decrease computational requirements drastically, but at present, CSP from first principles is still not feasible at scale: only a few hybrid-framework systems have been experimentally validated\cite{xu2023experimentally,ferey2005chromium}.

We propose a data-driven approach that is designed to enhance and supplement \textit{ab initio} methods in reticular-materials discovery. Owing to inherent rigidity and directionality of molecular building blocks, the atomic arrangement in MOFs is primarily determined by the connectivity of their constituent counterparts---metal-containing SBUs and organic linkers---i.e., by topology. Therefore, in terms of artificial intelligence, the CSP of MOFs can be defined as a classification task. In this work, as a source of training data, we employed the Quantum MOF (QMOF) database\cite{rosen2021machine}, containing a diverse collection of curated experimental structures. The initial set of 20,375 entities was preprocessed to decompose crystalline frameworks into organic and inorganic counterparts, identify underlying net topology, and eliminate duplicates (in terms of atomic arrangement and building blocks). Taking into account the data-hungry nature of neural networks, we excluded from consideration structures with infrequent topologies. The selected classes contain more than 50 instances and cover 80.6\% of correctly processed crystal graphs. It should be noted that the main reason for eliminating entities was some issues at decomposition and topology assignment stages (62.1\% of the initial set); 5,675 unique combinations of molecular building blocks and their corresponding topologies (according to O’Keeffe’s nomenclature\cite{o2008reticular}) were chosen for further training, validation, and testing. Consistently with previous findings\cite{alexandrov2011underlying}, the topology distribution proved to be highly uneven: 66.7\% of structures had one of three most common topologies (\textbf{pcu}, \textbf{sql}, and \textbf{rna}), and the total number of topologies was 19 (Figure 1A). Consequently, the distributions of topological features are also imbalanced and diverse. Underlying nets of most of examined MOFs (76.2\%) comprise one inequivalent node and one inequivalent edge, i.e., show transitivity [1 1], though five other transitivity types are present as well (Figure 1B). Connectivity of nodes expressed as the number of points of extension varies from (3)-c to (12)-c (Figure 1C), whereas 4- and 6-coordinated nets dominate (65.4\%). Finally, nearly two-thirds of the underlying nets are 3-periodic, while the rest are 2-periodic (Figure 1D).

The predictive task of CSP of MOFs cannot be carried out using atomic-level models owing to problem-specific input representation. Moreover, the incorporation of topology as an auxiliary descriptor into certain domain-specific neural networks\cite{yao2021inverse,cao2023moformer} makes them unsuitable for the problem in question. The recently developed coarse-grained crystal graph neural network\cite{korolev2024coarse} (CG\textsuperscript{2}-NN) appears to be the most appropriate architecture. The original CG\textsuperscript{2}-NN operates on a complete bipartite graph with nodes representing metal ions and organic linkers, and therefore single-atom nodes were substituted here with SBU ones to fully adapt the model to current needs (Figure S1). Elemental\cite{weston2019named} and molecular\cite{molfeat} embeddings were employed to describe inorganic and organic counterparts, respectively. By assigning symmetry-aware features to linkers, we incorporated domain knowledge into the model. Specifically, an extra neural network was trained to predict the number of maximal subgraphs (in a molecular graph) that can be moved into each other under the action of a graph automorphism (Figure S2); the last layer’s output served as a compact representation in addition to the general-purpose molecular embedding. The proposed descriptor strongly correlates with the number of points of extension (Figures S3 and S4), making topology prediction potentially easier by limiting the labels to the same connectivity type. After that, \textit{k}-fold cross-validation was performed on the modified CG\textsuperscript{2}-NN.

\begin{figure}
    \centering
    \includegraphics[width=17cm]{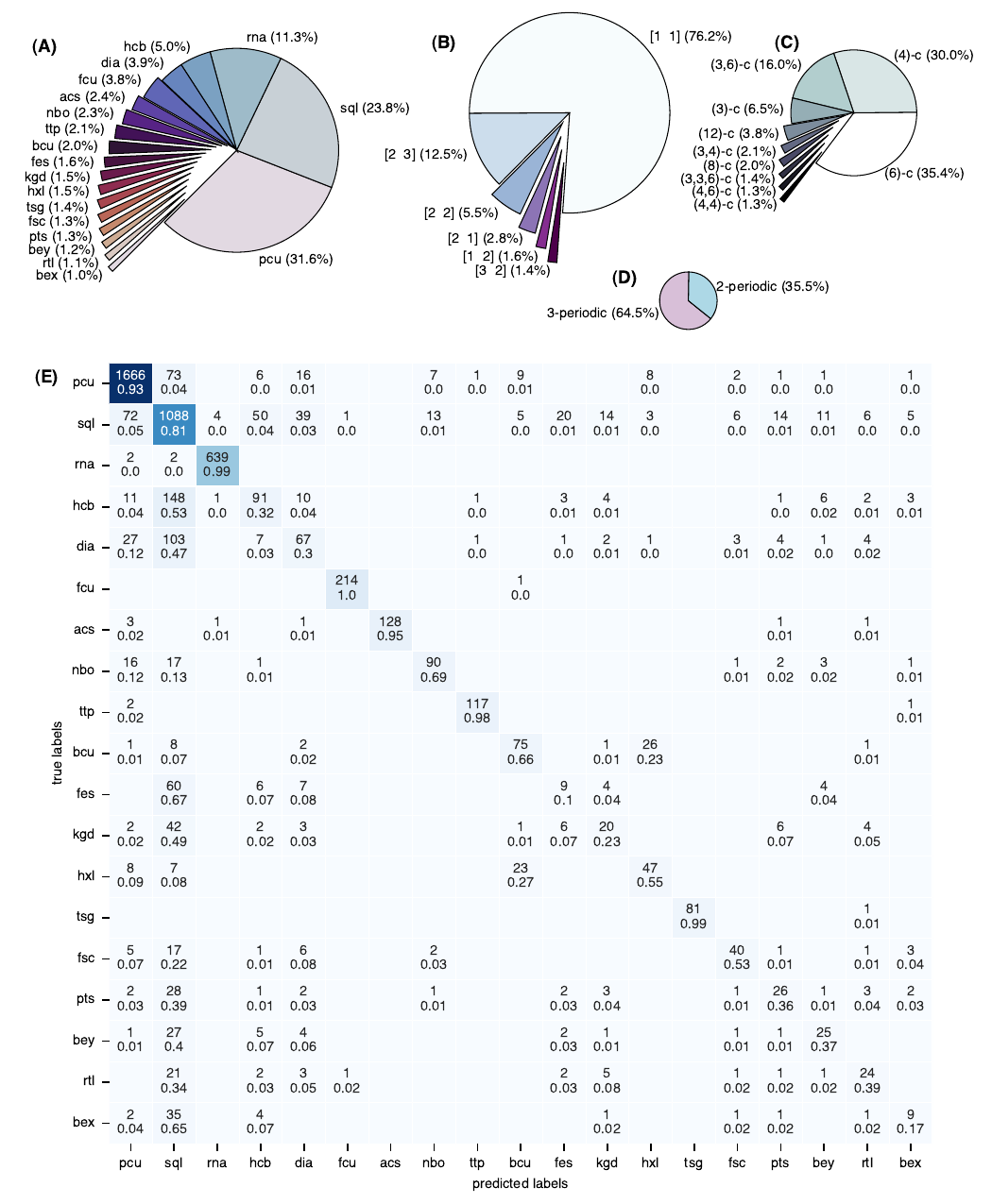}
    \caption{\textbf{An overview of the training dataset and model performance.} Distributions of (A) the underlying net topology, (B) transitivity, (C) connectivity, and (D) periodicity corresponding to a subset of the Quantum Metal-Organic Framework (Q-MOF) database. (E) The confusion matrix of the topology classification results obtained on the test dataset.}
    \label{fig:fig1}
\end{figure}

\begin{figure}
    \centering
    \includegraphics[width=17cm]{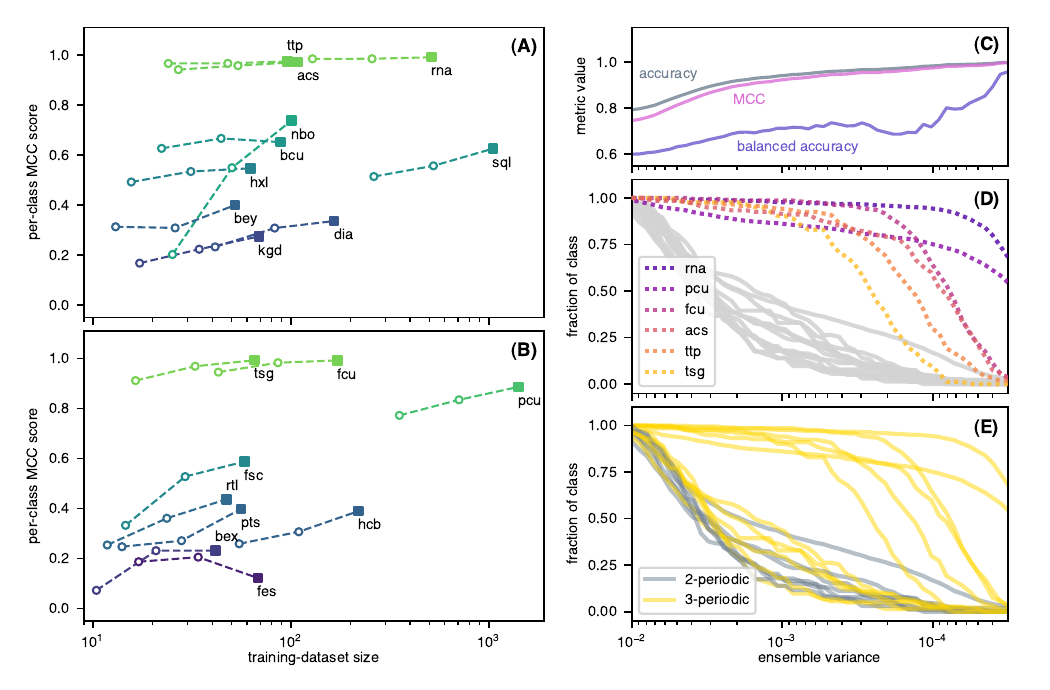}
    \caption{\textbf{Per-class and uncertainty-aware model performance.} (A, B) The per-class Matthews correlation coefficient (MCC) for classifying a topology as a function of training-dataset size. (C) The target classification metrics (overall accuracy, balanced accuracy, and MCC) as a function of model ensemble variance. (D, E) Proportions of topologies containing in-domain entities in accordance with model ensemble variance.}
    \label{fig:fig2}
\end{figure}

A confusion matrix provides an initial evaluation of algorithm performance (Figure 1E). The diagonal elements corresponding to the number of correctly identified topologies are the most numerous items in most rows and columns, indicating relatively high per-class recall and precision values, respectively. Cross-validated overall (balanced) accuracy determined on an external test set reached 0.785 ± 0.012 (0.595 ± 0.027, mean ± SD). The substantial difference in these metrics can be attributed to model biasing toward over-represented classes. Nevertheless, the similarity between the balanced accuracy value and the probability of a coin toss, i.e., performance of the random binary classifier, should not lead to misinterpretation. Via weighted sampling of labels on the basis of the observed topology distribution (Figure 1A), an overall (balanced) accuracy of 0.186 ± 0.021 (0.057 ± 0.007) could be achieved. Other important metrics—precision, recall, and F1 score—are 0.775 ± 0.012, 0.785 ± 0.012, and 0.772 ± 0.013, respectively; these parameters of a dummy classifier are 0.184 ± 0.018, 0.186 ± 0.021, and 0.184 ± 0.019. The complicated task of multiclass classification of imbalanced data can be further analyzed by considering the Matthews correlation coefficient (MCC): a robust alternative to accuracy and F1 score\cite{chicco2020advantages}. CG\textsuperscript{2}-NN yielded a decent MCC of 0.737 ± 0.015. The weighted random sampler, as expected, showed this metric’s value close to zero (0.008 ± 0.021). In addition, we assessed the model’s performance on an extended dataset that included classes with more than 10 entities (Figure S5). The metric’s values were lower than those in the case of the main dataset; overall accuracy, balanced accuracy, and MCC reached 0.752 ± 0.010, 0.517 ± 0.033, and 0.705 ± 0.012, respectively (Figure S6). At the same time, the CG\textsuperscript{2}-NN retained superb efficiency in the ranking of topologies by yielding a top-\textit{k} accuracy score of 0.835 ± 0.010 and 0.875 ± 0.010 for \textit{k} of two and three, respectively. Under this scenario, a minimalistic set of highly probable topologies should be used as a starting point for \textit{ab initio} CSP of MOFs.

Although the proposed data-driven approach is generally effective, its per-class performance ranges from excellent to unsatisfactory. It is worth noting that common topologies manifested higher values of the metric in comparison to under-represented ones (Figures 2A and 2B). For instance, MCC is 0.923 ± 0.009 (\textbf{pcu}, 1791 entities), 0.719 ± 0.025 (\textbf{sql}, 1351), 0.992 ± 0.007 (\textbf{rna}, 643), 0.39 ± 0.17 (\textbf{bey}, 67), 0.43 ± 0.11 (\textbf{rtl}, 61), and 0.22 ± 0.23 (\textbf{bex}, 54), and Spearman’s rank correlation coefficient between the number of topology occurrences and corresponding MCC reached 0.42 (Figure S7). For a more accurate assessment of model scalability, we trained CG\textsuperscript{2}-NNs on a smaller portion of data (25\% and 50\%), while retaining the original class distribution (Figures 2A and 2B). In most cases, per-class MCC substantially increased as the training-dataset size grew; a few exceptions were easily predictable (\textbf{acs}, \textbf{rna}, and \textbf{ttp}), as were minor (\textbf{fes}) topologies. Therefore, MOF CSP formulated as a predictive task is expectably sensitive to data scarcity, and this drawback can be overcome by supplementing the training set with thousands of novel experimentally resolved structures\cite{moghadam2017development}.

\begin{figure}
    \centering
    \includegraphics[width=17cm]{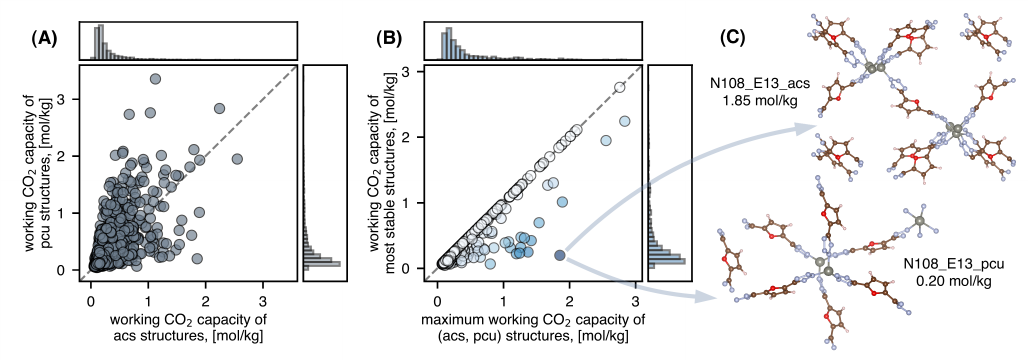}
    \caption{\textbf{High-throughput computational screening of hypothetical structures.} (A) A scatter plot of CO\textsubscript{2} working capacity values corresponding to polymorphs with topologies \textbf{acs} and \textbf{pcu}. (B) A scatter plot of CO\textsubscript{2} working capacity values corresponding to polymorphs with the highest performance measure and with the stablest topology. (C) Atomic structure of crystalline structures with the largest difference in CO\textsubscript{2} working capacity between the analyzed polymorphs.}
    \label{fig:fig3}
\end{figure}

A different way to improve model performance involves restricting the domain of applicability to subfields of chemical space that are exploited with low prediction error, i.e., epistemic uncertainty. To do so, the deep ensemble technique\cite{lakshminarayanan2017simple} is applied. In particular, structures are discarded if there is a significant difference in predictions made by a set of CG\textsuperscript{2}-NNs. As one can see in Figure 2C, narrowing the maximum variance of model outcomes leads to an increase in cumulative metrics. After exclusion of 26\% (31\%) of highly uncertain predictions, overall accuracy (MCC) exceeded 0.9. Improving balanced accuracy was more challenging: 71\% of entities needed to be eliminated to attain the same threshold value. The key to achieving better performance lies in prioritizing six specific labels (\textbf{acs}, \textbf{fcu}, \textbf{pcu}, \textbf{rna}, \textbf{tsg}, and \textbf{ttp}), as opposed to the less-consistent ensemble predictions of other classes (Figure 2D). The highlighted topologies are all 3-periodic (Figure 2E), and this property sets them apart from other well-represented classes, e.g., 2-periodic \textbf{hcb} and \textbf{sql}.

After assessing the accuracy of the new approach and exploring ways to enhance it, we can move on to practical applications. High-throughput computational screening of \textit{in silico}–generated compounds is a widely used method for discovering new functional MOFs. The field faces an ongoing problem of polymorphism, prompting the question of whether top-performing structures can be accessed synthetically. To illustrate how data-driven CSP can address this issue, we integrated the CG\textsuperscript{2}-NN with the workflow for generating MOFs suitable for flue gas separation\cite{bae2011development}. Adsorption efficiency of hypothetical frameworks (Figures S8–S11) reticulated from identical building blocks greatly differed judging by their topologies, as evidenced by grand canonical Monte Carlo calculations accompanied by machine learning (Figures 3A and S12). Specifically, mean absolute error and symmetric mean absolute percentage error between CO\textsubscript{2} working capacities of frameworks with topologies \textbf{acs} and \textbf{pcu} reached 0.16 mol/kg and 21\%, respectively. Therefore, the ranking of candidate materials for flue gas separation may be influenced by additional restrictions on polymorphs. Among the 529 pairs (\textbf{acs}, \textbf{pcu}) considered, the \textbf{pcu} polymorph has higher CO\textsubscript{2} working capacity than the corresponding \textbf{acs} framework in 361 cases (68\%). At the same time, the former topology was predicted to be the stablest in 491 systems (93\%). Not surprisingly, 100 top-ranking frameworks with and without analysis of only preferable topologies correlated at Spearman’s rank correlation coefficient of only 0.67 (Figure 3B). In this context, the framework consisting of the Zn\textsubscript{3}(tetrazole)\textsubscript{6} SBU and furan linker seems to be a representative example (Figure 3C): a polymorph with topology \textbf{acs} is in the top 2\% quantile of the CO2 working capacity distribution (1.85 mol/kg), whereas a stabler structure with the \textbf{pcu} topology ranks in the top 55\% quintile (0.20 mol/kg).

The CSP algorithm operating in a high-throughput manner is a missing puzzle piece of reticular-materials design. After the analysis of the given use case, it is evident that the other stages of the workflow are either well designed or actively being developed. Domain-specific crystalline constructors\cite{lee2021computational} make it possible to generate hypothetical MOFs \textit{in silico} at the scale of millions of instances\cite{kang2023multi}. The structures produced in this way are then typically optimized by fast-operating algorithms, including neural network interatomic potentials\cite{sriram2023open}. Furthermore, adsorption performance of resulting crystals is estimated by means of grand canonical Monte Carlo simulations\cite{dubbeldam2016raspa}; machine learning can be optionally utilized to assign atomic-level quantities, including atomic charges\cite{korolev2021parametrization}. Finally, top-ranked hypothetical structures are chosen as promising candidates, while their experimental validity is ignored in most cases. Relative stability of polymorphs in distinct systems can be determined by a combination of \textit{ab initio} CSP techniques and advanced finite-temperature calculations, but this methodology is not easily scalable for high-throughput applications. By contrast, the presented approach is aimed at reproducing the thermodynamic landscape of diverse chemical systems, similarly to experimentally resolved MOFs. The inherent weak ability of artificial intelligence algorithms to operate in an extrapolative way (beyond the applicability domain) can be overcome by extending a training dataset and quantifying model uncertainty.

\section{Conflicts of interest}
\label{sec:conflicts}
There are no conflicts of interest to declare.

\section{Data and code availability}
The source code accompanying this study was made publicly available at \url{https://doi.org/10.5281/zenodo.14052914}.

\newpage

\bibliographystyle{unsrt}
\bibliography{refs}

\end{document}